**Evaluating AI-Powered Learning Assistants in Engineering Higher Education: Student Engagement, Ethical Challenges, and Policy Implications**

**Ramteja Sajja**[1,*] (rsajja@tulane.edu)
**Yusuf Sermet**[1] (ysermet@tulane.edu)
**Brian Fodale**[2]
**Ibrahim Demir**[1,3] (idemir@tulane.edu)

[1] ByWater Institute, Tulane University, New Orleans, USA 70118
[2] Office of Teaching, Learning, and Technology, University of Iowa, Iowa City, USA 52242
[3] River-Coastal Science and Engineering, Tulane University, New Orleans, USA 70118
[*] Corresponding author, Email: Ramteja Sajja, rsajja@tulane.edu

**Abstract**
As generative AI becomes increasingly integrated into higher education, understanding how students engage with these technologies is essential for responsible adoption. This study evaluates the Educational AI Hub, an AI-powered learning framework, implemented in undergraduate civil and environmental engineering courses at a large R1 public university. Using a mixed-methods design combining pre- and post-surveys, system usage logs, and qualitative analysis of students' AI interactions, the research examines perceptions of trust, ethics, usability, and learning outcomes. Findings show that students valued the AI assistant for its accessibility and comfort, with nearly half reporting greater ease using it than seeking help from instructors or teaching assistants. The tool was most helpful for completing homework and understanding concepts, though views on its instructional quality were mixed. Ethical uncertainty, particularly around institutional policy and academic integrity, emerged as a key barrier to full engagement. Overall, students regarded AI as a supplement rather than a replacement for human instruction. The study highlights the importance of usability, ethical transparency, and faculty guidance in promoting meaningful AI engagement. A total of 71 students participated across two courses, generating over 600 AI interactions and 100 survey responses that provided both quantitative and contextual insights into learning engagement.

## 1. Introduction

Artificial Intelligence (AI) is increasingly being integrated into educational settings, reshaping traditional teaching and learning methodologies. From intelligent tutoring systems to automated grading tools, AI technologies are enabling more dynamic, efficient, and personalized learning experiences. These systems adapt to individual student needs, offering tailored feedback, real-time support, and enhanced engagement (Wangdi, 2024; Niveditha et al., 2023). At the same time, AI automates administrative tasks, such as scheduling and grading, allowing educators to focus on pedagogical goals (Murdan & Halkhoree, 2024). The potential of AI to transform education is significant, yet it also brings forth pressing challenges, including concerns around data privacy, algorithmic bias, and ethical use of AI in academic environments (Wangdi, 2024; Niveditha et al., 2023).

The integration of AI in education is not only enhancing instructional delivery but also transforming pedagogical models, curriculum design, and educator roles. AI-driven education is shifting from conventional, teacher-led approaches to more student-centered, interdisciplinary, and personalized frameworks (Kong & Yang, 2024; Siddiqui et al., 2025). Teachers are increasingly positioned as facilitators guiding self-directed learning processes and promoting metacognition and ethical reasoning (Almenara et al., 2024). Furthermore, pedagogical frameworks such as AI-TEACH and dual-contrast models are emerging to encourage analogical thinking, systems thinking, and ethical awareness in AI-integrated classrooms (Dai et al., 2023; Dai, 2024).

A central aspect of AI's value in education lies in its ability to enable personalized and adaptive learning. By leveraging machine learning algorithms and real-time analytics, AI-driven platforms such as intelligent tutoring systems can adapt instructional content based on student performance and learning behavior (Gyonyoru & Katona, 2024; Sari et al., 2024). These platforms have demonstrated improvements in academic outcomes, student engagement, and retention (Sari et al., 2024; Moleka, 2023). For example, the Artificial Intelligence-Enabled Intelligent Assistant (AIIA) framework supports adaptive learning by providing real-time quiz generation, flashcards, and personalized pathways using natural language processing (Sajja et al., 2024). Similarly, GPT-powered educational assistants have been shown to reduce barriers to engagement by offering flexible and discipline-specific support (Sajja et al., 2023a).

Moreover, AI technologies support the creation of adaptive learning paths, helping students navigate content at their own pace, identify knowledge gaps, and receive immediate feedback tailored to their specific needs (Hashem et al., 2024; Jian, 2023). This level of customization fosters deeper learning and increased motivation among students (Singh, 2024; Altinay et al., 2024). Conversational AI tools, such as those described in the Educational AI Hub framework, demonstrate how course-specific assistants can deliver accurate, context-aware answers, particularly in data-heavy disciplines like environmental science, public health and engineering (Zhang et al., 2023; Sermet and Demir, 2021). These systems also show promise in improving accessibility and inclusivity through enhanced document parsing and response generation (Sajja et al., 2025a).

AI applications in education are already diverse and rapidly expanding. AI chatbots serve as virtual tutors and administrative assistants, capable of answering student queries 24/7 and providing instructional support (Jain et al., 2024). Recommendation systems integrated into Learning Management Systems (LMS) help suggest appropriate resources, assignments, or next steps based on students' academic performance and interests (Kumar et al., 2025; Leong et al., 2024). Meanwhile, automated grading tools improve efficiency and consistency in assessment while providing timely feedback (Bah, 2024; Aravindh & Singh, 2024). Tools like the AI-powered Floodplain Manager (FPM) preparation assistant further demonstrate the applicability of AI (Pursnani et al., 2025) in vocational training, achieving high accuracy in adaptive quizzes and flashcards for certification readiness (Sajja et al., 2025b). Recent advancements also include fine-tuned domain-specific embedding models for educational question answering, optimized for semantic retrieval in academic documents such as course syllabi. These models significantly improve over existing baselines, supporting applications in academic chatbots, retrieval-augmented generation (RAG), and LMS integration (Sajja et al., 2025c). Similarly, the HydroLLM-Benchmark evaluates LLMs on hydrology-specific tasks, highlighting their potential for discipline-focused education (Kizilkaya et al., 2025).

In addition to these applications, AI-enhanced LMS platforms play a critical role in shaping student experience. Platforms that support personalized learning, adaptive assessments, and real-time feedback have been shown to improve academic performance and increase engagement (Alotaibi, 2024; Luo, 2023; Ikhsan et al., 2025). Students also report higher satisfaction when LMS systems are user-friendly and offer timely support features like conversational agents and analytics dashboards (Hamzah et al., 2025; Saqr et al., 2023). Furthermore, student self-efficacy has been identified as a mediator in the relationship between AI tool adoption and learning success (Gao, 2024). However, disparities in access, faculty readiness, and ethical concerns around data usage continue to pose implementation challenges (Alotaibi, 2024; Ikhsan et al., 2025).

In the context of higher education, AI is becoming especially relevant as institutions seek to improve student outcomes, optimize operations, and better prepare learners for a digital economy. Personalized learning experiences not only promote inclusivity but also enhance student satisfaction and achievement (Londoño, 2024; Deri et al., 2024). Administrative efficiencies, such as automated admissions, course scheduling, and performance analytics, allow institutions to operate more strategically (Dilmi & Sakri, 2024). AI is also being incorporated into curricula to help students acquire critical digital competencies needed in the evolving workforce (Vieira et al., 2022; Pursnani et al., 2023). Furthermore, learning analytics tools built on large language models like GPT-4 can provide real-time insight into students' cognitive and emotional engagement, supporting data-informed instructional strategies (Sajja et al., 2023b).

AI's role in fostering collaborative and interdisciplinary learning is also gaining momentum. Tools like virtual reality environments, peer-supported simulations, and AI-assisted group projects have demonstrated success in improving engagement and communication (Ramasamy, 2024; Msambwa et al., 2025). Curricula are increasingly incorporating AI literacy as a foundational skill, often starting at the K-12 level, to prepare students for ethical and effective use of AI (Dai, 2024; Ng et al., 2022; Yue et al., 2022). Models such as embodied or analogical pedagogies help demystify AI systems and encourage critical, cross-disciplinary thinking (Dai et al., 2023; Siddiqui et al., 2025).

While the benefits of personalized AI-supported learning are evident, they come with considerable limitations and ethical concerns. AI systems require extensive data collection, raising questions about data privacy and informed consent (Mimoudi, 2024). Bias in AI algorithms can lead to unequal treatment of students, especially those from underrepresented backgrounds (Santos et al., 2024). Furthermore, current AI models may not sufficiently support the development of students' independent learning or critical thinking skills (Cui & Alias, 2024). Technological infrastructure, teacher readiness, and resource availability also affect the success of AI integration (Santos et al., 2024). Over-reliance on automation risks depersonalizing education, making it vital to preserve the role of human educators in fostering intellectual and emotional growth (Mimoudi, 2024).

To navigate these challenges, institutions must adopt systems-thinking approaches that balance innovation with ethics, equity, and sustainability (Katsamakas et al., 2024). Faculty development, digital access policies, and evidence-based implementation strategies are essential to ensure successful and responsible AI integration across disciplines (Doğan et al., 2024; Zawacki-Richter et al., 2019).

Given this evolving landscape, it is critical to understand how students perceive, interact with, and are affected by AI-powered tools in real academic settings. This study explores the use of an AI-powered learning assistant in civil and environmental engineering courses, focusing on student experiences, usage patterns, and the impact on learning outcomes. By examining the interplay between trust, convenience, and engagement, as well as the ethical considerations of AI use, this research aims to provide insights that inform the responsible and effective integration of AI in higher education.

While numerous studies have explored AI's potential in education, few have empirically examined its role in discipline-specific engineering contexts where ethical trust and usability intersect. Grounded in sociotechnical and technology acceptance perspectives (e.g., UTAUT2 and human-centered learning frameworks), this study addresses that gap by investigating how students' perceptions of trust, ethics, and engagement shape the adoption of AI learning assistants in engineering education. This theoretical framing links the study to broader discussions of human-AI collaboration and responsible technology integration in higher education.

The study is guided by the following research questions (RQ): RQ1) What are the primary ethical concerns and trust issues students associate with using generative AI in education, particularly regarding academic integrity and content reliability? RQ2) How do students perceive the quality, convenience, and comfort of AI-generated learning support compared to traditional human assistance? RQ3) How does the perceived usefulness of AI tools for specific academic tasks, such as understanding concepts, completing assignments, and studying, affect students' acceptance and engagement? RQ4) What factors contribute to students' hesitation or apprehension about using AI in their education, and how do they envision the future role of AI in higher education?

## 2. Methodology

This study employed a mixed-methods approach to evaluate the use and impact of the Educational AI Hub, an AI-powered learning assistant designed to support civil and environmental engineering students. The methodology integrates quantitative and qualitative data sources to provide a comprehensive understanding of how students interacted with the tool, how they perceived its usefulness, and how it influenced their learning experiences. Key components of the methodology include a detailed description of the AI tool's features, participant demographics, survey instruments administered before and after tool usage, system-generated usage data, and qualitative analysis of the open-ended prompts and questions students posed to the AI chatbot. This multifaceted approach enabled both the measurement of behavioral patterns and the exploration of student perceptions within the context of real course settings.

### 2.1. Educational AI Hub Features

The Educational AI Hub (Sajja et. al., 2025a) is a comprehensive AI-powered learning assistant designed to support civil and environmental engineering students through a variety of intelligent, personalized learning tools. The system is integrated into the course learning environment and offers six core features aimed at enhancing student understanding, engagement, and academic performance. Below is a description of each functionality, which will be accompanied by corresponding screenshots in the final version.

**Notes Generation**: This feature allows students to input a topic of interest, upon which the system generates concise, structured summaries. The notes are synthesized from credible academic sources and are designed to help students quickly grasp key concepts. This supports efficient review and serves as a valuable study aid for self-paced learning. Figure 1 shows the interface for Notes Generation, illustrating how a student can enter a topic and receive AI-generated summarized content.

**AI-Assisted Chatbot**: The AI-assisted chatbot provides real-time responses to student queries related to course content. It retains a memory of the last three messages within a session, enabling it to maintain conversational context. Additionally, it personalizes its tone based on the student's interaction style, enhancing the sense of engagement and support. This function serves as an always-available virtual tutor, helping bridge communication gaps outside of classroom hours. As shown in Figure 2, the chatbot interface allows students to input course-related questions and receive contextual, adaptive replies in real time.

**Flashcards Generation**: By inputting a topic, students can automatically generate flashcards grounded in Bloom's Digital Taxonomy (Churches, 2010). The cards are designed to cover multiple levels of cognitive learning, from basic recall (e.g., definitions) to more advanced skills (e.g., application and analysis). This method helps diversify study approaches and deepens conceptual understanding. Figure 3 demonstrates the Flashcards Generation feature, displaying sample cards generated based on a specific academic topic.

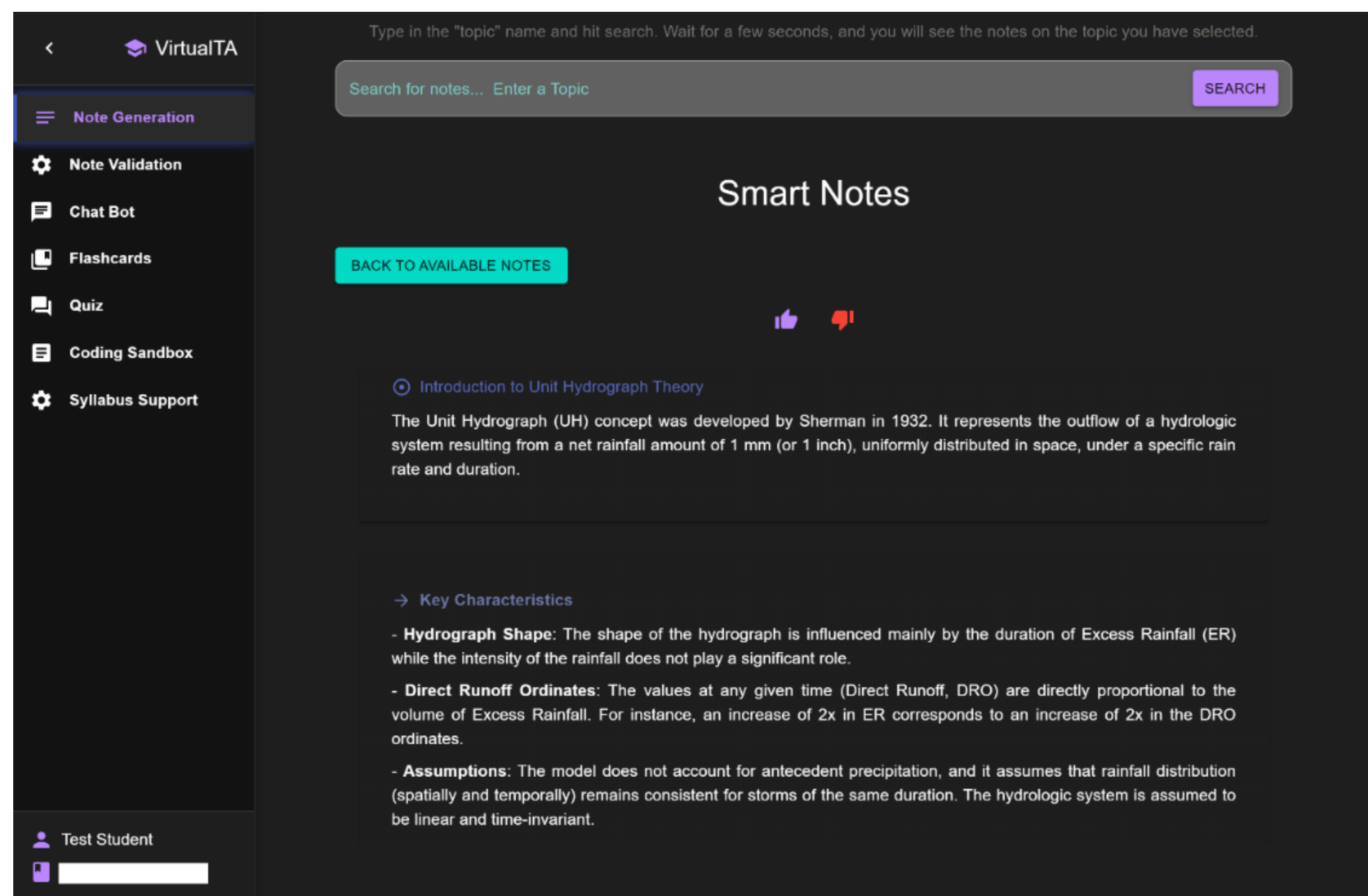

Figure 1. Interface for Notes Generation showing summarized content based on a user-provided topic.

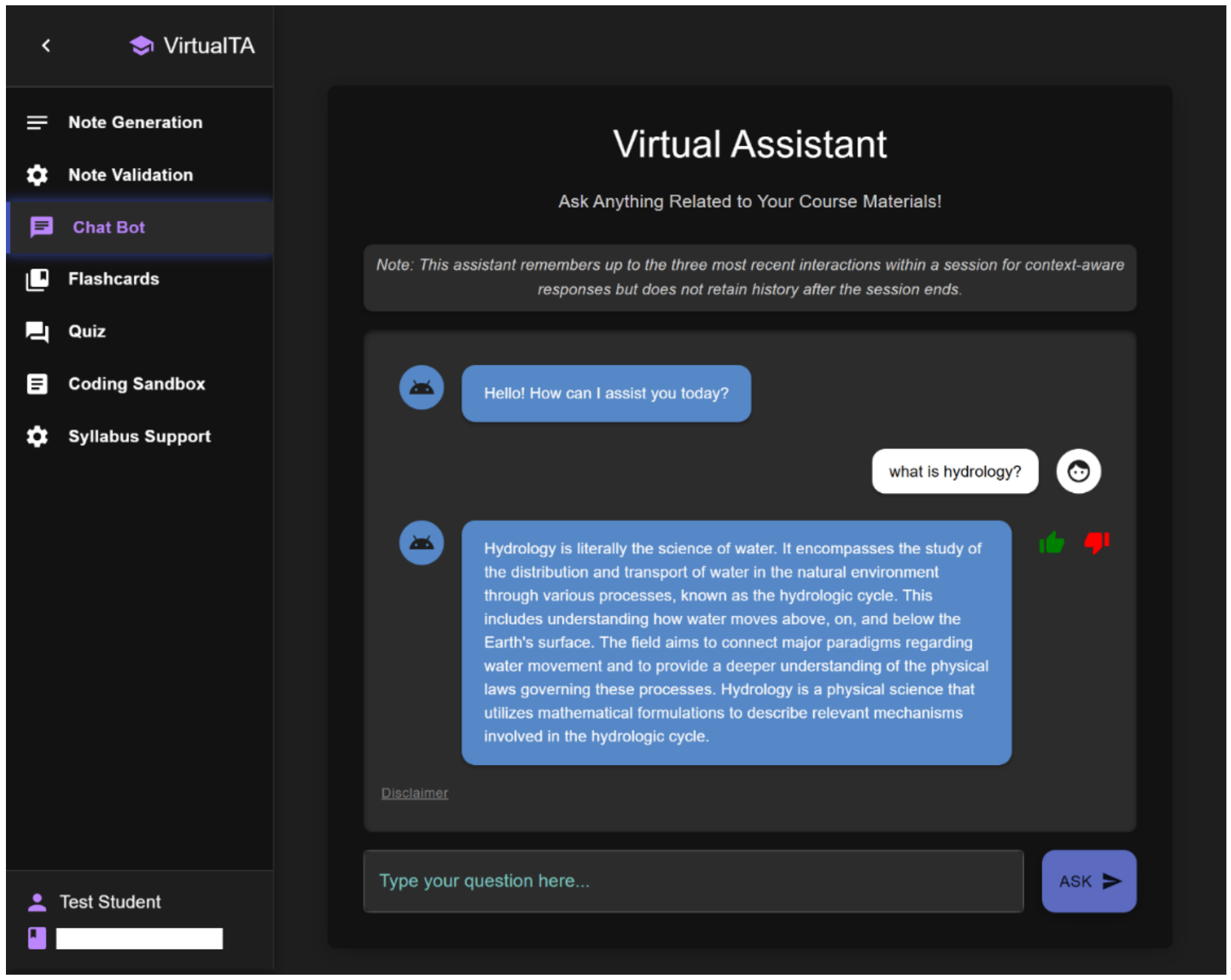

Figure 2. AI-Assisted Chatbot interface showing a sample interaction between a student and the system

**Quiz Generation and Grading**: Students can generate quizzes on any course-related topic to test their knowledge in a formative manner. The quizzes are also designed according to Bloom's digital Taxonomy (Churches, 2010), ensuring a mix of question types that assess both lower- and higher-order thinking skills. After completing the quiz, students receive instant feedback with explanations and a performance summary. They also have the option to download their quiz results, including responses and grades, for future study or reflection. Figure 4 presents the quiz interface, highlighting question generation, grading, and feedback functionalities. The system is designed to be flexible and intelligent in evaluating student responses. Student answers do not need to match the exact wording of the model-generated "ground truth" answer. Instead, the grading mechanism can interpret semantic similarity, allowing for variation in phrasing, synonyms, and structure. This flexibility supports a more authentic assessment of understanding and encourages students to express knowledge in their own words, an important aspect of higher order thinking as emphasized in Bloom's digital taxonomy. This nuanced grading approach is a distinct strength of the system and helps ensure fairness and meaningful feedback.

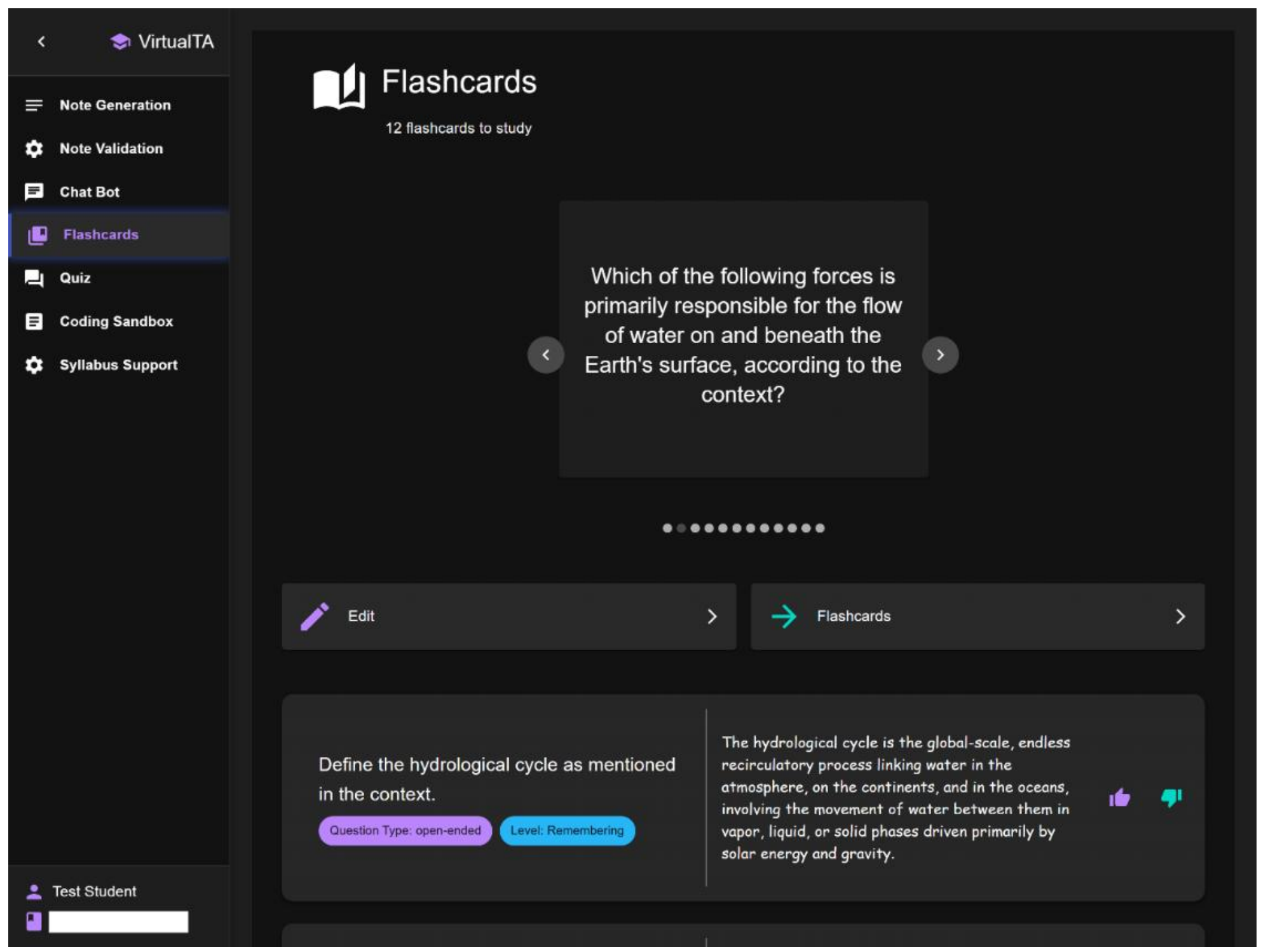

Figure 3. Flashcards Generation interface illustrating Bloom's Taxonomy-based cards

**Coding Sandbox:** For programming-related support, the Coding Sandbox allows students to submit coding questions and receive tiered assistance. The system first offers pseudocode to encourage independent problem-solving. If students need further help, it provides step-by-step code explanations, including executable code snippets. This functionality enhances problem-solving skills and supports students learning computational tools and languages used in engineering.

**Syllabus Support**: This feature enables students to ask administrative or logistical questions related to the course, such as exam dates, assignment deadlines, and office hours. The system retrieves relevant information from the syllabus and provides quick, accurate responses, reducing student uncertainty and improving course navigation.

## 2.2. Study Design

This study was conducted in two undergraduate courses within the Civil and Environmental Engineering department at a large R1 public university in the Midwest: one sophomore-level course and one senior-level course. These courses were selected because they provided diverse contexts in which students regularly engage with analytical, design, and problem-solving tasks, making them suitable settings to explore the use of AI-powered learning tools in engineering education. While data were collected from both courses, the analysis does not distinguish between them, as the sample sizes were not balanced and the study's aim was to understand general patterns of engagement and perception across different academic levels rather than to make direct comparisons.

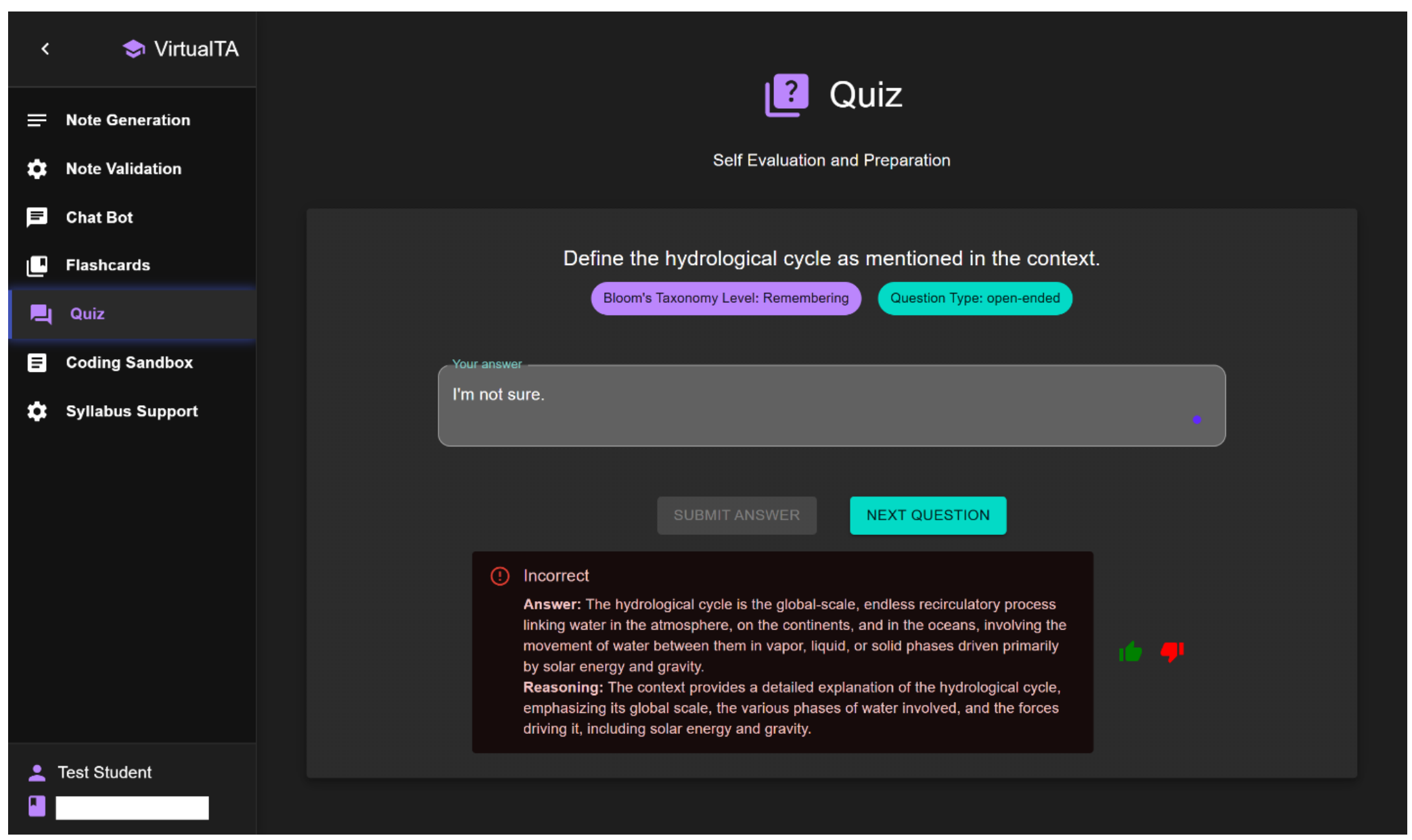

Figure 4. Quiz Generation interface with real-time grading and feedback capabilities

A total of 77 students were enrolled across both courses (60 in the sophomore-level course and 17 in the senior-level course). Of these, 65 students voluntarily participated in the study, yielding an overall participation rate of 84 percent. Participation across survey phases was as follows: 18 students completed only the pre-survey, 44 students completed both the pre- and post-surveys, and 3 students completed only the post-survey. Among the participants, 25 identified as female and 40 as male. The majority were Bachelor of Science in Engineering students (n = 62), while three were enrolled in a Master of Science program. The sample also included ten honors students and fifty-five non-honors students, providing representation across academic levels and achievement tracks.

Students were introduced to the Educational AI Hub during class and given access through the course learning management system. Of the 65 participants, 48 actively used the AI Hub at least once. Within individual courses, 35 of 60 students (58 percent) in the sophomore-level course and 13 of 17 students (76 percent) in the senior-level course engaged with the tool.

To evaluate the tool's effectiveness, the study used a mixed-methods approach combining: (i) Pre- and post-usage surveys to assess changes in student perceptions, trust, ethical concerns, and learning outcomes; (ii) System-generated usage logs to track feature interaction patterns, timing, and behavioral trends; (iii) Qualitative analysis of student-generated questions submitted to the AI system, which offered insights into students' learning needs, conceptual challenges, and engagement strategies during tool use.

This research followed an explanatory sequential mixed-methods approach in which quantitative data from surveys and usage logs were analyzed first, followed by qualitative examination of student-AI interactions to contextualize and interpret the patterns observed. Integration occurred during interpretation, quantitative results guided the development of qualitative coding themes, enabling a comprehensive understanding of how students engaged with the AI system, how they perceived its value, and how these perceptions aligned with observed behavioral evidence.

### 2.3. Data Collection & Analysis

**Pre- Usage Survey**: The pre-usage survey was administered at the beginning of the semester to establish a baseline understanding of students' familiarity with and attitudes toward generative AI. It focused on capturing students' prior experiences using AI tools both for academic and non-academic purposes, their confidence in using such tools responsibly, and their expectations of AI's impact on learning, grades, and productivity. The survey also addressed students' trust in AI-generated responses and ethical concerns, including perceptions of academic misconduct and fairness in AI use among peers. Responses were collected using Likert-scale items, multiple-choice questions, and a few open-ended prompts. These baseline responses were later compared with post-usage survey results to assess changes in perceptions and attitudes following direct engagement with the Educational AI Hub.

**Post-Usage Survey**: The post-usage survey was administered at the end of the semester to evaluate students' experiences with the Educational AI Hub after sustained interaction. It included items measuring: (i) Perceived value and usability of the tool (e.g., quality of help, convenience of access, and comfort compared to human support like TAs and instructors); (ii) Perceived impact on learning activities, such as studying, understanding course concepts, completing homework, and overall academic performance; (iii) Trust and apprehension related to AI usage and confidence in effectively using the tool to support learning; (iv) Perceptions of academic integrity and ethical use of generative AI in coursework; (v) Awareness of institutional AI policies and attitudes toward AI use restrictions in academic settings.

The survey included both Likert-scale questions and optional open-ended items, enabling students to elaborate on their opinions and provide feedback based on their actual tool usage.

**System Usage Data:** System-generated logs were collected throughout the semester to capture detailed records of how students interacted with the Educational AI Hub. This quantitative data provided objective measures of engagement and behavioral trends, including: (i) feature-specific usage frequency, such as how often students used the chatbot, notes generator, flashcards, quizzes, and coding sandbox; (ii) query types and content patterns, reflecting the nature of students' questions or prompts; (iii) session-level data, including timestamps, session duration, and frequency of repeat use; and (iv) interaction depth, such as whether students downloaded quizzes, generated study materials, or engaged in follow-up interactions These logs enabled a granular view of tool engagement across both courses and were used to correlate usage behaviors with survey-reported outcomes, such as perceived helpfulness and learning gains.

In addition to quantitative usage metrics (e.g., number of quizzes generated, frequency of use), qualitative data were drawn from the questions students posed to the AI system. These questions, submitted as part of quiz generation or exploratory use, provided insight into students' thought processes, misconceptions, and learning goals. To better understand how learners engaged with the tool, this student-generated content was analyzed and categorized into five types: Software Help, Conceptual / Theory Help, Assignment / Class Help, Miscellaneous / Irrelevant, and Technical Issues. This categorization helped identify patterns in how students leveraged AI to support their learning and the kinds of cognitive or practical challenges they sought to address.

**Open Ended Questions**: At the end of the post-usage survey, students were invited to respond to an optional open-ended question: "*If you would like, please share any final comments you have related to your experiences with generative AI in this course.*" While these responses were not included in the current analysis, the question was designed to offer space for additional reflections beyond structured survey items. For qualitative insights in this study, we instead focus on analyzing the actual questions students posed to the AI system during usage, as these provide direct evidence of how students engaged with the tool in context.

**Participant Recruitment and Ethics**: Students were recruited to participate in the study through announcements posted on the course LMS at the beginning of the semester. To encourage participation, students were offered a small amount of extra credit for completing the pre- and post-usage surveys. Participation was voluntary, and students who preferred not to engage with the AI tools or surveys were given the option to complete an alternative reflective essay for equivalent credit. The study protocol, including recruitment and data collection procedures, was reviewed and approved by the university's Institutional Review Board (IRB), ensuring compliance with ethical standards for research involving human subjects.

**Statistical Design and Power**: The quantitative analysis followed a repeated-measures pre–post design using both independent and paired comparisons to examine changes in student perceptions of trust, ethics, and usability before and after interaction with the Educational AI Hub. Descriptive and correlational analyses were used to explore associations among attitudes, trust, and engagement metrics.

A post hoc power analysis was conducted to assess the adequacy of the sample size (N = 65). Assuming a medium effect size (Cohen's $d = 0.5$) and $\alpha = 0.05$, the achieved statistical power for paired-sample tests exceeded 0.80, indicating sufficient sensitivity to detect meaningful pre–post differences in student perceptions. For correlational analyses, the achieved power for moderate associations ($\rho \approx 0.30$) was approximately 0.75, confirming that the study had adequate power to detect effects commonly observed in educational research (Cohen, 1988; Faul et al., 2007; Lakens, 2013). Given these parameters, the study design and sample size were considered statistically sufficient to support valid interpretation of observed trends across both quantitative and qualitative components.

## 3. Results

This section presents the results of the study, drawing on quantitative and qualitative data to examine how students engaged with the Educational AI Hub and how they perceived its value in supporting their learning. Findings are organized around key themes aligned with the research questions, including student perceptions of AI versus traditional human support, concerns related to trust and academic integrity, perceived usefulness across academic tasks, usage patterns across tool features, and the relationship between engagement and learning outcomes. Together, these results offer insight into both the benefits and limitations of integrating generative AI tools into undergraduate engineering education.

### 3.1. Student Perceptions of AI vs. Human Assistance

To assess how students perceived the Educational AI Hub in comparison to traditional forms of support (e.g., instructors, teaching assistants, or peers), the post-usage survey included items evaluating three dimensions: quality of help, convenience of access, and comfort in seeking help. Responses were compared on a 5-point scale ranging from "Much Worse" to "Much Better," with summary data visualized in Figures 5.

As shown in Figure 5, student ratings indicated that AI-generated help was generally perceived as comparable but not superior to traditional human assistance in terms of instructional quality. When asked to evaluate the quality of help received from the Educational AI Hub, 34% of students rated it as *worse* than human support, while 49% felt it was *about the same*. Only 17% of students perceived the AI-generated help as *better* than that provided by instructors, TAs, or peers. In contrast, ratings for convenience were more favorable toward the AI tool. A majority of students (68%) reported that the Educational AI Hub was *more convenient* to access compared to human assistance. Meanwhile, 21% rated it as *equally convenient*, and 11% found it to be *less convenient*.

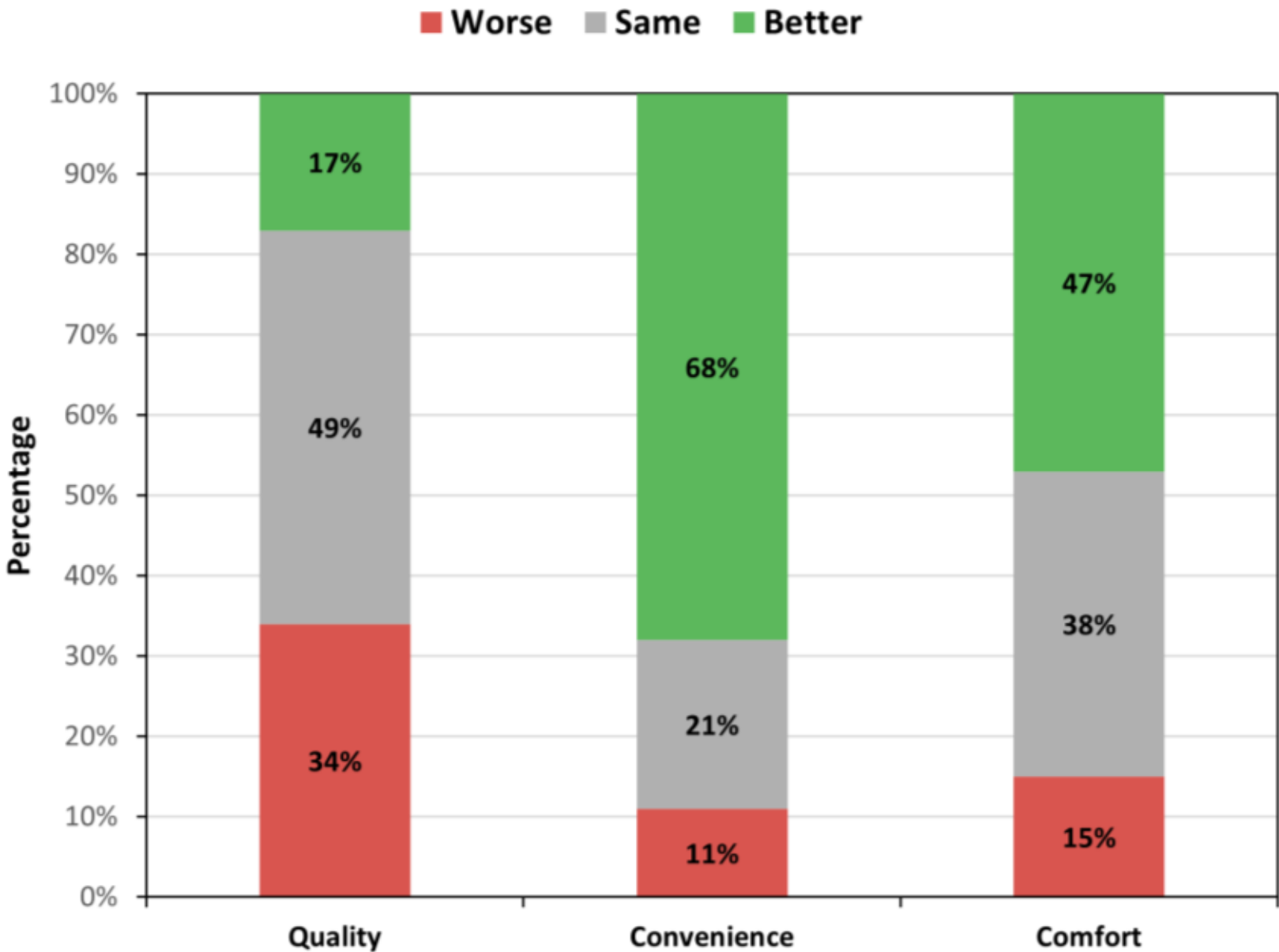

Figure 5. Student ratings of AI Hub vs. human assistance by quality, convenience, and comfort ($N = 47$)

Student perceptions regarding comfort level in seeking help leaned positive. Nearly half of the respondents (47%) indicated that they felt *more comfortable* using the AI Hub than asking for help from human resources. An additional 38% reported *no difference* in comfort level, and only 15% felt *less comfortable* using the AI-based system.

### 3.2. Challenges and Barriers to Use of AI

While students generally found the AI Hub comfortable and convenient, survey results also revealed several concerns and barriers related to its use. As shown in Figure 6, post-survey responses captured student perceptions on trust, academic integrity, and apprehension regarding AI. When asked whether they distrusted the AI Hub's responses, 42% of students disagreed, indicating trust in the tool, while 27% agreed and 31% remained neutral. This suggests that although a majority did not distrust the tool, a notable portion of students were either unsure or skeptical of its reliability.

In terms of general apprehension about using AI in the course, student responses were more polarized: 49% agreed they felt apprehensive, while 24% disagreed, and 27% reported neutral feelings. This highlights a potential emotional or psychological barrier, where even students who used the tool may have had reservations about relying on it fully. Concerns regarding academic integrity were also prominent. When asked whether they believed the use of AI in coursework undermines academic integrity, 27% agreed, 24% were neutral, and 49% disagreed. While nearly half of the students dismissed the concern, the remaining half expressed some level of uncertainty or agreement, indicating ethical ambiguity surrounding AI-assisted learning.

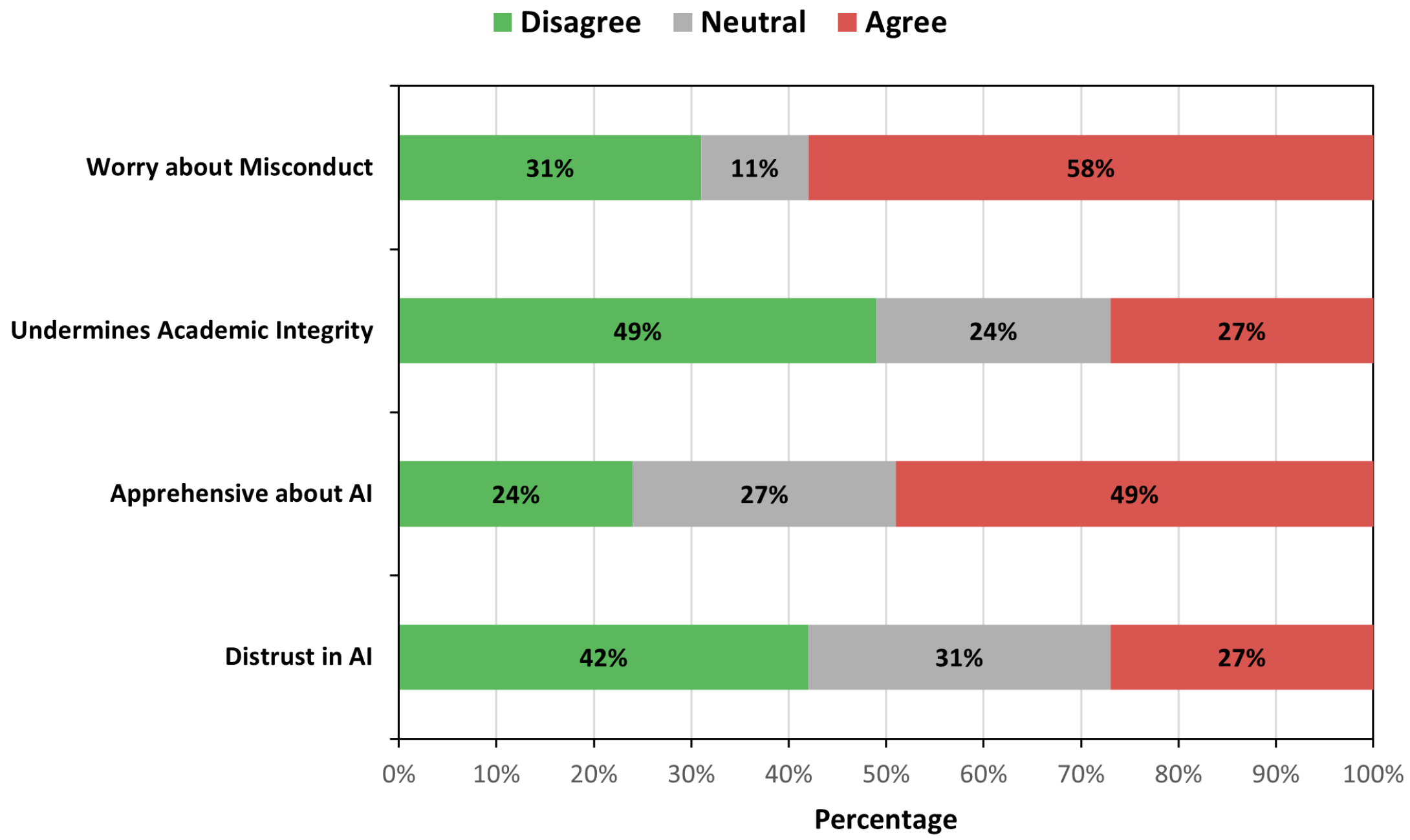

Figure 6. Student concerns about AI accuracy, academic integrity, and potential misconduct (N = 45)

Finally, a majority of students (58%) reported worry about being accused of academic misconduct when using AI, compared to 31% who disagreed, and 11% who were neutral. This concern about potential policy violations or misunderstandings may serve as a significant deterrent to full engagement with AI tools, regardless of their utility. These findings suggest that even when students recognize the functional value of AI in education, emotional, ethical, and institutional factors, such as trust, concerns about academic integrity, and uncertainty around misconduct, can influence their engagement. These aspects are important to consider for the effective design, communication, and implementation of AI tools in educational settings.

### 3.3. Perceived Helpfulness Across Learning Activities

Students were asked to rate the perceived usefulness of the Educational AI Hub across various learning dimensions, including understanding course concepts, studying, completing homework, improving grades, and supporting overall learning. The results, shown in Figure 7, highlight generally favorable attitudes toward the AI tool's learning support capabilities, particularly in concept acquisition and homework assistance.

A majority of students (61%) agreed that the AI Hub helped them learn course concepts effectively, while only 22% disagreed and 17% were neutral. Similarly, 63% of students agreed that it helped them complete homework, with 20% disagreeing and 17% remaining neutral. These two areas, concept learning and homework, emerged as the strongest perceived benefits of the AI assistant.

Perceptions of usefulness for studying and overall learning were also generally positive, though slightly more mixed. One-third (33%) of students agreed that the AI Hub helped with studying, while 33% disagreed and 35% remained neutral. For overall learning, 59% of students reported a positive impact, 17% disagreed, and 24% were neutral.

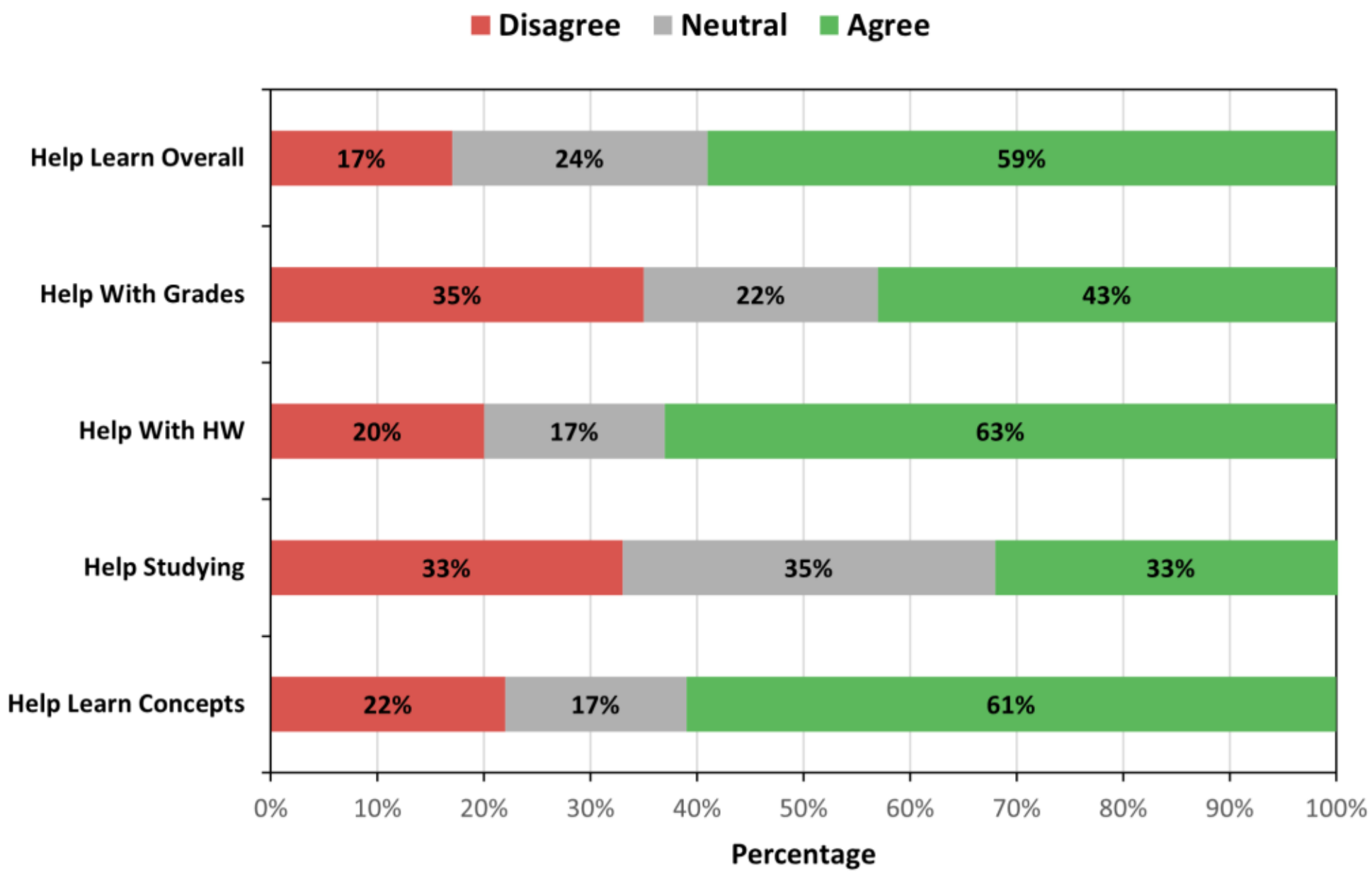

Figure 7. Student perceptions of AI's helpfulness across different learning activities (N = 46)

In contrast, the perceived helpfulness in improving grades was more divided: only 43% agreed the tool helped them earn a higher grade, while 35% disagreed and 22% remained neutral. This suggests that while students valued the AI Hub for learning support and task completion, they were more cautious in attributing performance outcomes directly to its use. These results suggest that students saw the AI Hub as a valuable support tool for comprehension and task completion, though they were more ambivalent about its effect on performance metrics like grades. The positive perception of its role in learning aligns with earlier findings on usage patterns and comfort levels, reinforcing its role as a complementary, not primary, learning aid.

### 3.4. Usage Patterns and Most Frequently Used Features

System log data revealed varied usage patterns across the different components of the AI tool. As shown in Table 1, chatbot messages were by far the most frequently used feature within the Educational AI Hub, with 555 total interactions and usage by 46 out of 48 students who engaged with the tool at least once (96%). This indicates that the chatbot function was the primary component of the Educational AI Hub that students interacted with. However, it is important to note that these logs do not capture students' use of external AI tools such as ChatGPT and thus cannot speak to their overall AI usage beyond the Hub.

In contrast, activity-based features, including quizzes, flashcards, and notes, were used much less frequently. These features accounted for 75 interactions across 15 unique students, representing only 31% of users. As detailed in Table 2, the most accessed structured feature was flashcard review, used by 17% of students. Other moderately used features included notes generation and review (each at 8%), and flashcard and quiz generation (each at 6%). Features like quiz completion and quiz review were used by only 2–4% of students, suggesting these were explored by a very small subset.

Descriptive statistics in Table 3 further support these patterns. Students averaged 13.12 total AI interactions (SD = 11.81), most of which came from chatbot use (Mean = 11.56). In contrast, engagement with structured activity features was substantially lower, with an average of 1.56 interactions per student. Although not shown in the tables, system logs also indicated that AI tool usage tended to spike around assignment deadlines and exam periods, suggesting students turned to the system primarily for last-minute review and clarification.

Table 1. Interaction types and their usage by students

| Interaction Type | Frequency | Unique Students | Unique Percentage |
|---|---|---|---|
| Activity | 75 | 15 | 31% |
| Chatbot Message | 555 | 46 | 96% |

Table 2. Breakdown by Activity Type and their usage by students

| Activity Type | Frequency | Unique Students | Unique Percentage |
|---|---|---|---|
| Flashcard Generation | 5 | 3 | 6% |
| Flashcard Review | 10 | 8 | 17% |
| Notes Generation | 5 | 4 | 8% |
| Notes Review | 5 | 4 | 8% |
| Quiz Answered | 39 | 2 | 4% |
| Quiz Completed | 4 | 1 | 2% |
| Quiz Generation | 4 | 3 | 6% |
| Quiz Review | 3 | 2 | 4% |

Table 3. Summary statistics for activities and usage by students

| Variable | N | Mean | Std. Dev. | Min | Pctl. 25 | Median | Pctl. 75 | Max |
|---|---|---|---|---|---|---|---|---|
| AI Interactions | 48 | 13.12 | 11.81 | 1 | 5.75 | 9.5 | 17 | 62 |
| Total Chatbot Messages | 48 | 11.56 | 9.437 | 0 | 5 | 9 | 17 | 41 |
| Total Activities | 48 | 1.562 | 6.687 | 0 | 0 | 0 | 1 | 45 |

### 3.5. Query Topics and Keyword Insights

To better understand student engagement with the Educational AI Hub, all chatbot messages were systematically analyzed and categorized into thematic topics using a combination of manual coding and automatic keyword-based classification. As presented in Table 4, the majority of student inquiries (40.4%) fell under the *Software Help* category. These questions primarily involved the use of technical platforms such as ArcGIS and MATLAB, and included troubleshooting issues, performing specific tasks, and understanding workflow steps within these tools.

Table 4. Questions categories and their distribution

| Question Category | N | Percentage |
|---|---|---|
| Software Help | 234 | 40.4% |
| Conceptual / Theory Help | 176 | 30.4% |
| Assignment / Class Help | 113 | 19.5% |
| Miscellaneous / Irrelevant | 38 | 6.6% |
| Technical Issues | 18 | 3.1% |

The second most prevalent category was Conceptual / Theory Help, accounting for 30.4% of the questions. These queries were primarily related to understanding core civil engineering and hydrology concepts. Students asked for definitions, theoretical explanations, and clarification of formulas or models, such as the limitations of Snyder's unit hydrograph or the meaning of saturation excess.

Assignment / Class Help comprised 19.5% of the interactions. Questions in this category included those seeking clarity on homework content, assignment scope, and course logistics such as instructor contact information or TA assignments. A smaller proportion of queries (6.6%) were categorized as Miscellaneous / Irrelevant, capturing low-effort, off-topic, or ambiguous inputs. These included overly simple questions like “What is 1+1?” as well as social niceties like “Thank you!” that did not require academic intervention.

Lastly, Technical Issues represented 3.1% of student messages. These questions dealt with system errors, display bugs, or failures within software environments (e.g., receiving null values or blank viewports), indicating technical barriers that occasionally disrupted learning. Table 5 provides representative examples of each category to illustrate the breadth and nature of student interaction with the AI system. Together, this analysis highlights the tool’s relevance in addressing both instructional and technological support needs within the course environment.

### 3.6. Interaction Type and Learning Satisfaction

To understand the factors influencing student engagement with the AI-powered learning tool, Spearman correlations were calculated between post-survey variables and actual system usage and the top five correlations were considered. As summarized in Table 6, the results highlight the significant role of trust, comfort, and perceived usefulness in shaping how frequently students interacted with the tool.

Table 5. Sample student questions by category

| Category | Example Question |
|---|---|
| Software Help | How do I convert an ArcGIS layer into a File Geodatabase feature class in ArcGIS Pro? |
| | How do I create a map layout of a Clip layer? |
| Conceptual / Theory Help | What are some limitations of the Snyder's unit hydrograph method? |
| | What is the definition of saturation excess? |
| Assignment / Class Help | What does Assignment 3 cover? |
| | What is my TA's name and UIowa email address? |
| Miscellaneous / Irrelevant | What is the answer to the following equation? 1+1=? |
| | Thank you! |
| Technical Issues | I keep getting null for all the table values |
| | Why is my viewport blank in civil 3d? |

The strongest association was a negative correlation between distrust in AI and system usage ($\rho = -0.38$, $p < 0.01$), indicating that students who lacked trust in the AI assistant were significantly less likely to use it. This emphasizes the importance of trust as a critical enabler of engagement. Conversely, comfort level with the AI (relative to human assistance) was positively associated with usage ($\rho = 0.31$, $p < 0.05$), suggesting that students who felt more at ease using the tool engaged with it more frequently. Similarly, students who believed the AI assistant was helpful for completing homework were more likely to use it ($\rho = 0.30$, $p < 0.05$). Although not statistically significant, perceived convenience compared to human assistance ($\rho = 0.22$) and perceived helpfulness for improving grades ($\rho = 0.14$) also showed positive correlations with usage, indicating a general trend where more favorable attitudes aligned with higher levels of engagement.

### 3.7. Student Perceptions of Ethics, Policy, and Future Use of AI in Education

In addition to usage patterns and engagement factors, students' perceptions of the ethical dimensions of AI in coursework and their preferences for future integration offer critical context for understanding adoption and sustainability. Survey responses revealed generally positive attitudes toward the ethical use of AI and a desire for thoughtful but supportive institutional policies.

Table 6. Spearman correlations between survey variables and AI tool usage (N = 44)

| Survey Variable | Spearman's ρ | Significance |
|---|---|---|

| Distrust in AI | –0.38 | $p < 0.01$ |
|---|---|---|
| Comfort Level with AI | 0.31 | $p < 0.05$ |
| Helpfulness for Completing Homework | 0.30 | $p < 0.05$ |
| Perceived Convenience | 0.22 | ns |
| Helpfulness for Improving Grades | 0.14 | ns |
| *Note: ns = not significant* | | |

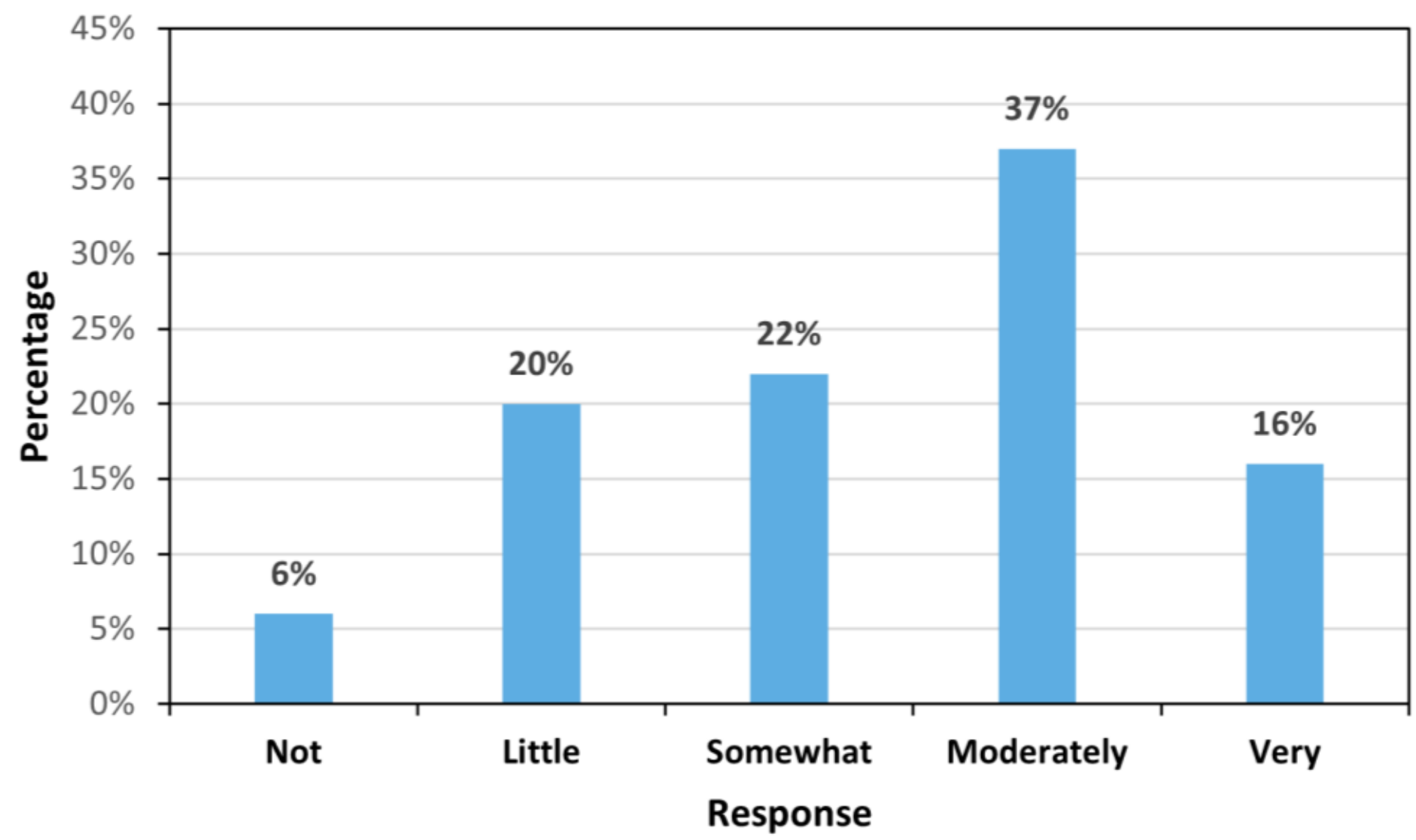

Figure 8. Student certainty about what constitutes academic misconduct when using generative AI (N = 51)

As shown in Figure 8, students expressed varying levels of certainty in identifying what constitutes academic misconduct when using generative AI. Specifically, 37% of respondents reported feeling “moderately” certain, 16% felt “very” certain, and 6% indicated they were “not at all” certain. However, a substantial portion (42%) remained uncertain or only somewhat confident, indicating that more clarity and guidance is needed from instructors and institutions to support responsible AI use.

When asked about how restrictive course policies should be regarding AI use, the majority of students favored flexibility over enforcement. As shown in Figure 9, 59% preferred “somewhat” restrictive policies, while another 33% favored even fewer restrictions. Notably, none of the students supported banning AI tools outright, suggesting a broad consensus in favor of inclusion rather than limitation.

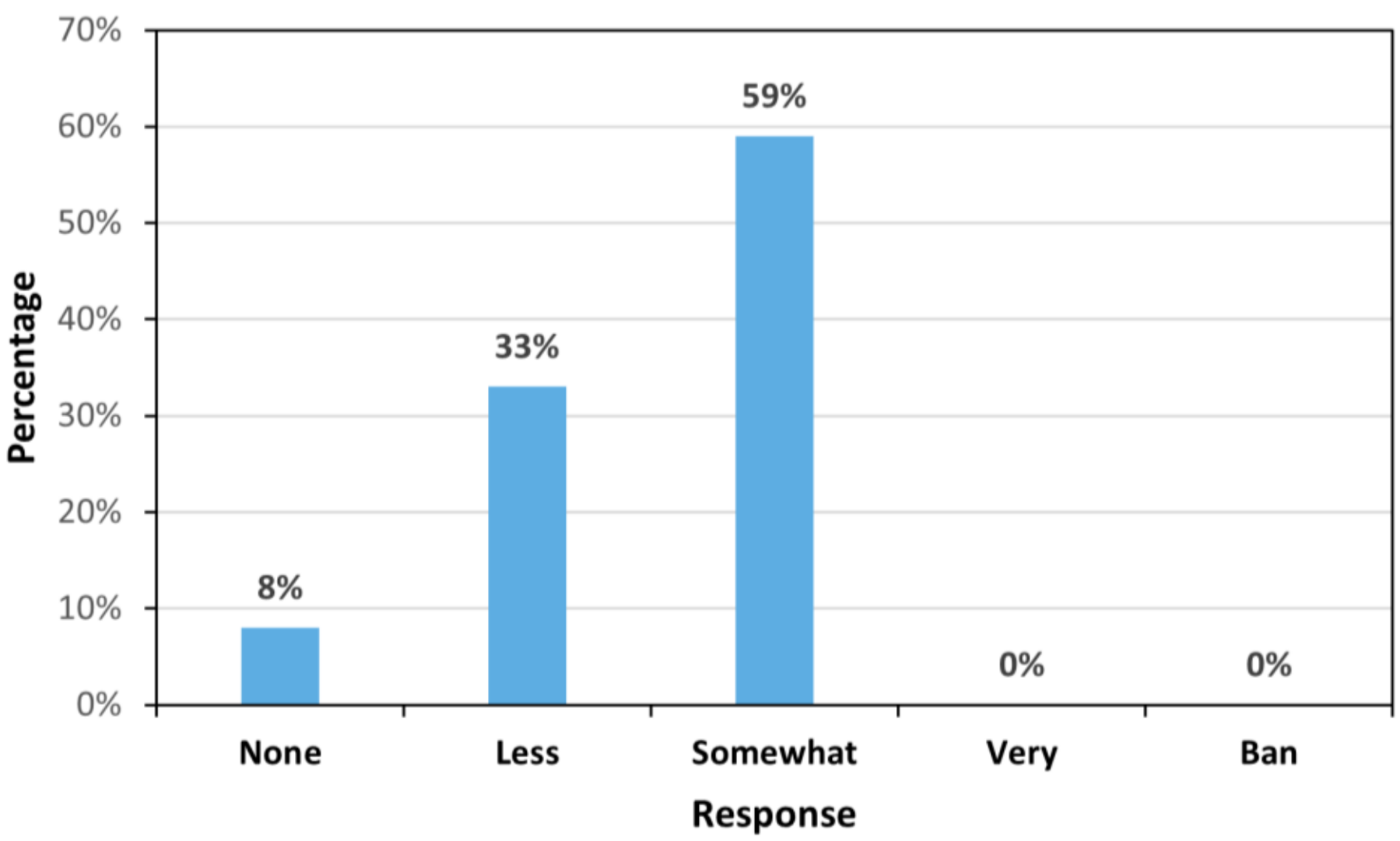

Figure 9. Student preferences for course policy restrictiveness regarding AI tool use (N = 51)

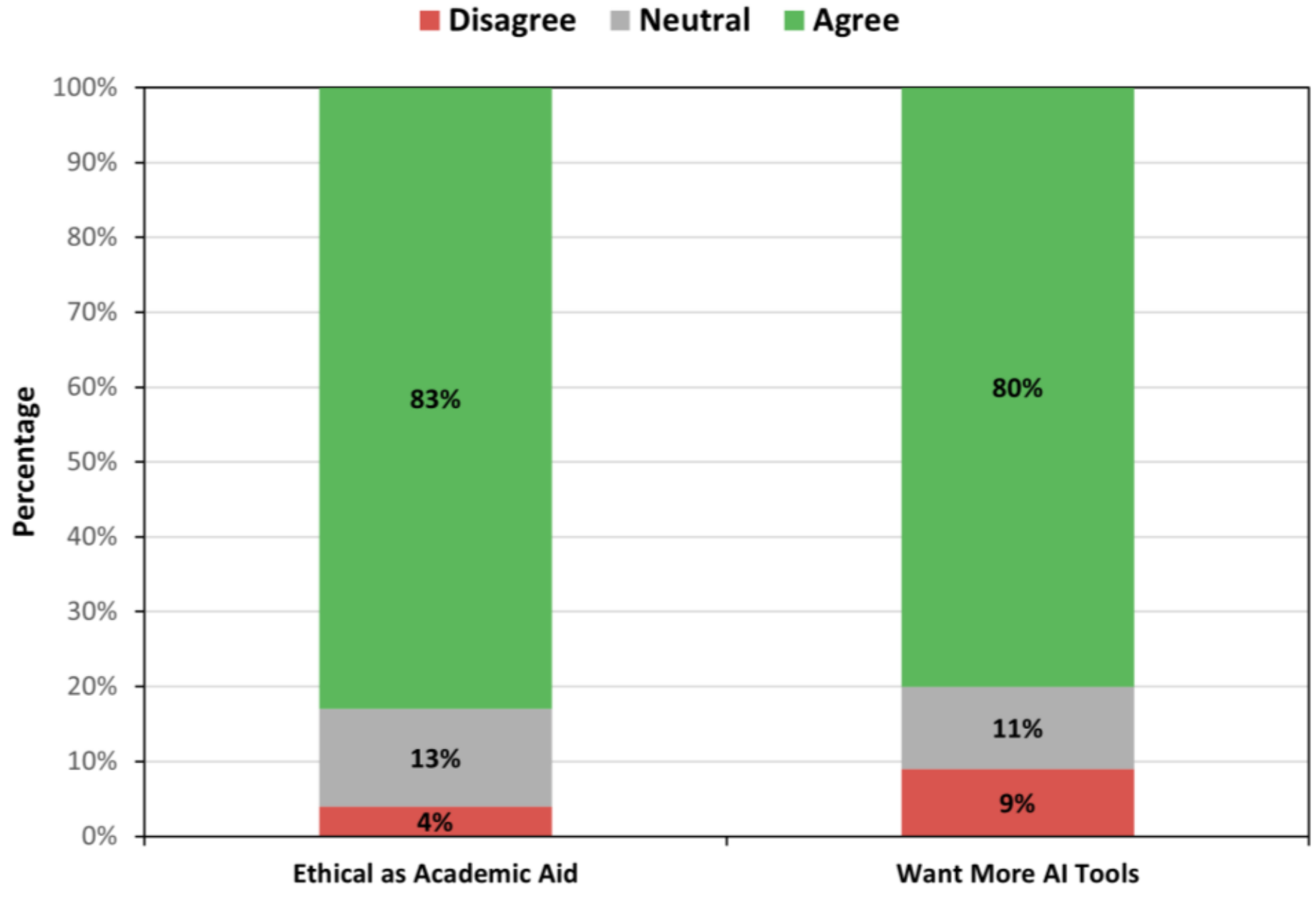

Figure 10. Student agreement on the ethical use of AI as an academic aid and interest in future integration of AI tools (N = 46)

These attitudes are reinforced by students' views on ethical legitimacy and future potential of AI in education. As illustrated in Figure 10, a large majority (83%) agreed that using AI as an academic aid is ethical, and 80% expressed interest in having similar tools available in other courses. These findings demonstrate strong student endorsement of AI when framed as a learning enhancer rather than a shortcut.

Together, these results suggest that while students are enthusiastic about AI tools and view them as ethically acceptable, institutional responsibility lies in ensuring transparency, setting clear expectations, and fostering informed use. Students appear to favor a balanced approach that promotes innovation without undermining academic integrity.

## 4. Discussions

Referring to research question RQ1, which asked about the ethical concerns and trust issues students associate with generative AI in education, the findings reveal both openness and significant hesitation. While students broadly viewed AI as ethically permissible, 83% agreed it is ethical to use AI as an academic aid, their uncertainty about institutional boundaries was striking. A majority (58%) reported concern about being accused of academic misconduct when using AI, and 42% indicated they were uncertain or only somewhat confident in identifying what constitutes a violation. This ambiguity suggests that institutions and instructors have not yet fully equipped students to navigate the ethical landscape of generative AI. This aligns with recent studies emphasizing the need for clearer institutional AI governance and literacy frameworks (Almenara et al., 2024; Doğan et al., 2024). Although most students trusted the AI tool's intentions, they lacked trust in the institutional systems surrounding it. This gap highlights an urgent need for course-level AI usage guidelines and clear communication from faculty about what constitutes acceptable use. Without such clarity, students are left to make ethical decisions in a gray area, which can dampen engagement and lead to anxiety rather than innovation.

Referring to RQ2, which examined how students perceive AI-generated support in comparison to human assistance, the results point to a crucial divide between instructional quality and usability. While only 17% rated the AI tool as superior in instructional quality, a majority (68%) found it more convenient, and nearly half (47%) felt more comfortable using it. This preference for AI convenience and comfort over traditional human support is especially revealing. It suggests that students may feel less intimidated, less judged, or more in control when interacting with an AI system, factors that can be particularly important in technical or high-pressure environments like engineering education. The findings raise important pedagogical questions about how instructors and TAs create spaces for help-seeking. If students turn to AI for its emotional neutrality and accessibility, it may point to a broader need for more approachable, responsive human support that matches AI's flexibility while offering superior instructional depth.

Referring to RQ3, which explored the relationship between perceived usefulness of AI tools and student engagement, the data underscore a clear link between task-specific relevance and tool adoption. Students who found the AI assistant helpful for completing homework or who felt comfortable using it were significantly more likely to engage with it. More broadly, students valued the AI most when it supported concrete academic outcomes, 63% said it helped with homework and 61% said it improved their understanding of course concepts. Similar task-oriented adoption patterns have been reported in adaptive learning contexts (Sari et al., 2024; Katona
& Gyonyoru, 2024). Engagement was less enthusiastic when it came to more abstract benefits like grade improvement (43%) or studying (33%). These patterns were reinforced by a qualitative analysis of the types of questions students posed to the AI chatbot. The majority of queries fell into three dominant categories: Software Help, Conceptual/Theory Help, and Assignment/Class Help. These categories closely align with the areas where students reported the most benefit from the tool, particularly in clarifying course content and completing assignments. This convergence of quantitative and qualitative evidence suggests that students are more inclined to use AI tools when their value is immediate, specific, and clearly tied to core academic tasks. This insight can guide the future design of AI tools and course integration, pointing to the importance of task-aligned features, scaffolding, and clear communication about how AI use can enhance, not just supplement, learning.

Referring to RQ4, which investigated factors behind students' hesitation to use AI and their visions for its future in higher education, the study revealed a student body that is cautiously optimistic. Nearly half of students (49%) reported feeling apprehensive about using AI in their course, and many expressed concerns tied to unclear expectations around academic integrity. However, their policy preferences clearly leaned toward openness and inclusion: 59% favored "somewhat restrictive" AI policies and 33% preferred even fewer restrictions, with no students supporting a total ban. These preferences suggest students do not want to be left without guidance, but they also reject heavy-handed restrictions. Notably, 80% of students indicated they would like to see similar AI tools used in other courses, a response tied to the Educational AI Hub specifically, but one that likely reflects a broader interest in AI-supported learning. Taken together, these results suggest that students are open to future AI integration, provided it is accompanied by thoughtful design, clear institutional support, and well-communicated ethical guidelines. In this context, policy clarity, not policy severity, is what students are asking for.

Taken together, the findings across all four research questions suggest that students are curious, cautious, and highly motivated to integrate AI into their learning, provided the right support structures are in place. They see clear value in AI tools for convenience and targeted academic support, but their engagement is shaped by factors such as trust, ethical clarity, and usability. When these conditions are met, students are more likely to experiment with and benefit from AI. However, when policies are vague or institutional support is inconsistent, even highly motivated learners may hesitate. As institutions move toward deeper AI integration in higher education, these findings highlight three essential priorities: designing tools that address specific academic challenges, ensuring that both faculty and students understand the ethical boundaries of use, and establishing flexible yet clearly communicated institutional policies. Together, these steps can ensure that AI adoption not only enhances learning but also fosters student confidence and upholds academic integrity.

## 5. Conclusion

This study examined how undergraduate civil and environmental engineering students engaged with an AI-powered learning assistant, focusing on students' perceptions of trust, ethical use, learning value, and usability. The findings reveal a generally positive outlook on the role of generative AI in education, with students appreciating the tool's accessibility, convenience, and ability to support specific academic tasks, particularly homework completion and conceptual understanding. Nearly half of the students reported feeling more comfortable seeking help from the AI assistant than from instructors or TAs, and most found it easier to access. However, while the tool was well-received in terms of usability, it was not universally seen as superior in instructional quality. Students remained cautious about its accuracy and expressed persistent concerns about academic integrity and institutional policy clarity.

These findings contribute to the broader understanding of AI integration in education by emphasizing that successful adoption is not solely a matter of functionality or technical design. Instead, the impact of AI depends heavily on how well it aligns with students' learning needs and how confidently students can engage with it in an ethically supported environment. Students do not view AI as a replacement for human support, but rather as a valuable supplement, especially when traditional resources are inaccessible or intimidating. The results also show that trust, perceived relevance, and emotional comfort are key predictors of student engagement with AI tools.

The implications for wider implementation are clear. Institutions looking to scale AI use in higher education must consider more than just access, they must provide clear, course-level policies that define acceptable AI use and reduce uncertainty around academic misconduct. Faculty plays a crucial role in this process and should be supported in helping students navigate AI-integrated coursework through explicit guidance, open discussion, and pedagogical alignment. Furthermore, AI tools themselves should be designed with context-awareness, memory retention, and course-specific alignment to ensure they meet students' real academic needs. The emphasis should be on empowering students to use AI responsibly and effectively, rather than leaving them to navigate ambiguous expectations.

Looking forward, future work should explore how AI tools impact long-term learning behaviors, self-regulation, and academic outcomes across diverse disciplines and student populations. More research is also needed to develop ethical AI literacy frameworks that help students critically evaluate and appropriately use AI technologies in their academic work. Beyond immediate perceptions, the study underscores the importance of AI literacy and ethical transparency as foundational competencies for future engineers. Embedding explicit AI policy discussions within coursework can strengthen students' trust and self-regulation, supporting sustainable integration of generative AI in STEM education. As AI becomes increasingly embedded in higher education, the goal must be not only to improve efficiency or access, but to foster equitable, trustworthy, and learner-centered environments. With thoughtful integration, clear institutional support, and ongoing dialogue, AI has the potential to enhance, not just digitize, the educational experience.

**Acknowledgements**
The authors would like to thank Jae-Eun Russell, Salim George, and Anna Smith for their invaluable support throughout the study. Their assistance with IRB approvals, coordination of study logistics, and thoughtful feedback during both the research and writing phases was instrumental to the success of this project. We also gratefully acknowledge the Office of Teaching, Learning, and Technology (OTLT) at the University of Iowa for their guidance and support in implementing the Educational AI Hub and facilitating this evaluation.

**Funding**
Funding for this project was provided by the University of Iowa's Innovations in Teaching with Technology Awards.

**Ethics Approval and Consent to Participate**
This study was approved by the Human Subjects Office Institutional Review Board (IRB), approval number 201707769. Informed consent was obtained from all participants prior to the administration of both pre- and post-surveys. Participation was entirely voluntary, and all individuals were fully informed about the purpose and procedures of the study. The authors affirm that this work was conducted in accordance with established ethical standards.